\documentclass[useAMS,usenatbib]{mn2e}

\usepackage{epsfig}

\def\prb{Pr\,{\sc iii}}
\def\ndb{Nd\,{\sc iii}}
\def\pra{Pr\,{\sc ii}}
\def\nda{Nd\,{\sc ii}}
\def\fea{Fe\,{\sc ii}}
\def\ya{Y\,{\sc ii}}
\def\ha{H$\alpha$}

\title[Pulsations of the roAp star HD~24712]
{Pulsations in the atmosphere of the
roAp star HD~24712 II. Theoretical models}

\author[H. Saio, T. Ryabchikova, and M.Sachkov]
{Hideyuki Saio$^{(1)}$\thanks{saio@astr.tohoku.ac.jp}, 
Tanya Ryabchikova$^{(2)}$, and Mikhail Sachkov$^{(2)}$\\
$^{(1)}$Astronomical Institute, Graduate School of Science, Tohoku University,
Sendai, 980-8578, Japan\\
$^{(2)}$Institute of Astronomy, Russian Academy of Science, Pyatnitskaya 48,
119017 Moscow, Russia 
}

\begin{document}

\date{}

\pagerange{\pageref{firstpage}--\pageref{lastpage}} \pubyear{2009}

\maketitle

\label{firstpage}

\begin{abstract}
We discuss pulsations of the rapidly oscillating Ap (roAp)
star HD~24712 (HR~1217) based on nonadiabatic analyses 
taking into account the effect of dipole magnetic fields. 
We have found that all the pulsation modes
appropriate for HD~24712 are damped; i.e., 
the kappa-mechanism excitation in the hydrogen ionization layers
is not strong enough to excite high-order p-modes with periods consistent 
with observed ones, all of which are found to be above the acoustic 
cut-off frequencies of our models.

The main (2.721~mHz) and the highest (2.806~mHz) frequencies
are matched with modified $l=2$ and $l=3$ modes, respectively.
The large frequency separation ($\approx~68~\mu$Hz) is reproduced 
by models which lay within the error box of HD~24712 on the HR diagram.
The nearly equally spaced frequencies of  HD~24712 indicate 
the small frequency separation to be as small as $~\approx~0.5~\mu$Hz. 
However, the small separation derived from 
theoretical $l=1$ and 2 modes are found to be larger
than $\sim3~\mu$Hz.
The problem of equal spacings could be resolved by assuming
that the spacings correspond to pairs of $l=2$ and $l=0$ modes; 
this is possible because magnetic fields significantly modify
the frequencies of $l=0$ modes.
The amplitude distribution on the stellar surface is strongly 
affected by the magnetic
field resulting in the predominant concentration at the polar regions.
The modified amplitude distribution of a quasi-quadrapole
mode predicts a rotational amplitude modulation consistent
with the observed one.

Amplitudes and phases of radial-velocity variations
for various spectral lines are converted to relations of
amplitude/phase versus optical depth in the atmosphere.
Oscillation phase delays gradually outward in the outermost layers
indicating the presence of waves propagating outward.
The phase changes steeply around $\log\tau\sim-3.5$, which supports
a $T-\tau$ relation having a small temperature inversion there. 
\end{abstract}


\section{Introduction}\label{intro}
HD~24712 (HR~1217, DO Eri) is a prototype of the rapidly oscillating Ap (roAp)
stars that consist of  about 40 members. The oscillations, with periods  
ranging from $\sim6$ to $\sim20$~min, are high radial-order p-modes
affected by strong magnetic fields of  1~kG to 25~kG.

The 6.15 minute light variation of HD~24712 was discovered by \citet{ku81}.
\citet{ma88} first discovered radial velocity (RV) variations with an amplitude
of $400\pm50$\,m\,s$^{-1}$. 
A WET (Whole Earth Telescope) campaign found 
eight frequencies with rotational side-lobes as presented in \citet{ku05};
the paper also gives a thorough review of preceding research on this star.   
From accurate RV measurements \citet{mkr05} obtained two additional frequencies.
\citet{pap1} (Paper I) studied oscillation amplitudes and phases corresponding 
to the two highest amplitude modes for $\approx$600 unblended spectral lines
of different elements/ions, and found a diversity of these pulsational 
characteristics in two thirds of them. 

The magnetic field of HD~24712 has been studied by many authors 
as reviewed in \citet{ry97}.
\citet{bag95} derived a polar magnetic strength of $B_p=3.9$~kG,
a rotational inclination (angle between line-of-sight and rotation axis)
of $i=137^\circ$ (or $43^\circ$) and a magnetic obliquity (the angle between 
the rotation and magnetic axes) of $\beta = 150^\circ$ (or $30^\circ$).  
\citet{ry97} found that a slightly higher strength of $B_p=4.4$~kG 
is more consistent with their polarimetric observations.
Recently, by inverting rotationally modulated polarized spectra, \citet{lu08} 
found that HD~24712 has a nearly dipole magnetic field with the strength 
varying between 2.2~kG and 4.4~kG, depending on the rotation phase.  

One of the reasons why oscillations of roAp stars are important lies in  
the fact that many (often regularly spaced) oscillation frequencies are 
excited simultaneously giving great potential for asteroseismic studies. 
One difficulty in applying asteroseismology to roAp stars is  that
strong magnetic fields affect the high-order p-modes in complex ways.
Although the frequency of an oscillation mode generally increases with
increasing magnetic field strength, it occasionally jumps down
by $\sim10$ to $30\,\mu$Hz and then starts increasing again \citep{cg00,sg04}. 
At present, there are three independent methods to calculate magnetic effects 
on high-order p-modes: the method of \citet{cg00} is based on a variational 
principle, the other two use truncated expansions with spherical harmonics
to present the angular dependences of eigenfunctions \citep{dg96,big00,sg04}.
The results from those three methods roughly agree with each other
as discussed by \citet{cu06} and by \citet{sa08}.

A further complexity arises if the rotation effect is taken into account.
\citet{bd02} investigated the oscillation properties taking into
account both effects of rotation and magnetic field.
In our present investigation, however, we disregard the effect of rotation on
the frequency and eigenfunctions for simplicity, hoping that the effect
is small because the pulsation periods, $\sim 6$~min, are very much
shorter than the $\approx12$-d rotation period of HD~24712.
 
Previously, theoretical pulsation models based on the method of \citet{sa05} 
were compared with the observed frequencies for $\gamma$ Equ \citep{gru08}, 
10 Aql \citep{hub08} and HD 101065 \citep{mkr08}.
Those attempts were more-or-less successful, although the required strength
of the magnetic fields in models tended to be larger than that measured 
by the Zeeman splittings of spectral lines.
In this paper we provide the same modelling of HD~24712 and for the first 
time we compare not only oscillation frequencies but also
the amplitude and phase variations in the atmosphere of HD~24712.

\section{Models}\label{models}
\subsection{Unperturbed models}

\begin{table}
\caption{Unperturbed model parameters}
\label{tab:models}
\begin{tabular}{@{}lcccccc}
\hline
Name & $M/{\rm M}_\odot$ & $X$ & $Z$ & $T(\tau)$ & He-dep. & conv \\
\hline
AD160 & 1.60 & 0.70 & 0.02 & Ap & Y & N  \\
AD165 & 1.65 & 0.70 & 0.02 & Ap & Y & N  \\
AD170 & 1.70 & 0.70 & 0.02 & Ap & Y & N  \\ 
AH165 & 1.65 & 0.70 & 0.02 & Ap & N & N  \\
AH165C & 1.65 & 0.70 & 0.02 & Ap & N & Y  \\
AD170Z25 & 1.70 & 0.695 & 0.025 & Ap & Y & N  \\
AH170Z25 & 1.70 & 0.695 & 0.025 & Ap & N & N  \\
AD150Z1 & 1.50 & 0.71 & 0.01 & Ap & Y & N  \\
AH150Z1 & 1.50 & 0.71 & 0.01 & Ap & N & N  \\
SD160 & 1.60 & 0.70 & 0.02 & SS & Y & N  \\
SD165 & 1.65 & 0.70 & 0.02 & SS & Y & N  \\
SD170 & 1.70 & 0.70 & 0.02 & SS & Y & N  \\
SH165 & 1.65 & 0.70 & 0.02 & SS & N & N  \\
\hline
\end{tabular}
\end{table}

For unperturbed models, we adopted spherically symmetric evolutionary
models calculated with OPAL opacity tables \citep{opal95}.
Table~\ref{tab:models} lists various assumptions adopted for each
series of evolutionary models.
Except for AH165C, envelope convection is suppressed,
assuming a strong magnetic field to stabilize convection.
For the envelope convection included in AH165C we used a local mixing-length 
theory with a mixing-length of 1.5 times the pressure scale height.

\begin{figure}
\begin{center}
\epsfig{file=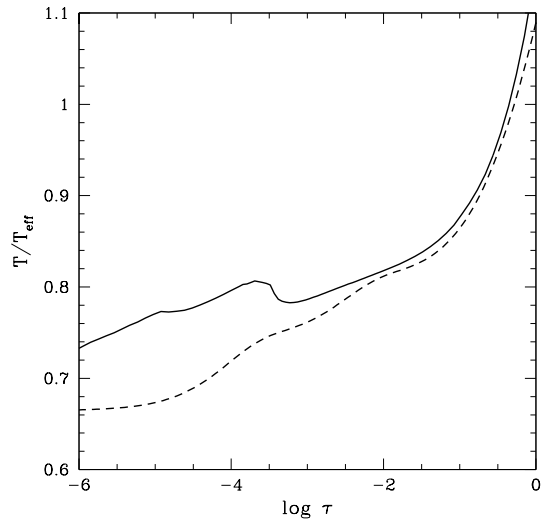,width=0.45\textwidth}
\end{center}
\caption{Two types of  $T-\tau$ relations employed.
The standard relation (dashed line) is obtained from eq.~(10)
of \citet{ss85}, while the Ap star relation (solid line) is adopted 
from \citet{sh09}.
}
\label{fig:ttau}
\end{figure}

We considered two types of $T-\tau$ relations in the optically thin outer
layers (Fig.~\ref{fig:ttau}). One of them is a standard relation
(denoted as ``SS'' in Table~\ref{tab:models}) 
obtained from eq.~(10) of \citet{ss85}.
The other one (denoted as ``Ap'' in Table~\ref{tab:models}) is a relation 
obtained by \citet{sh09} for the self-consistent atmospheric 
model of HD~24712 that takes into account element stratification 
including rare-earth elements.
The latter relation has a small temperature inversion at 
$\log\tau_{5000}\approx -3.5$ where abundances of praseodymium (Pr) and 
neodymium (Nd) increase steeply outward. 
(Although the optical depth in Shulyak et al.'s 
relation refers to
$\tau_{5000}$, optical depth at a wavelength of $5000$~\AA, 
we assume in our present paper that $\tau_{5000}$ is
not very different from Rosseland-mean optical depth $\tau$.)

We have computed models with and without helium depletion in the outermost
layers (the 6th column of Table~\ref{tab:models}). 
In the helium depleted models, the helium abundance is 
calculated as $Y = 0.01 + (Y_i-0.01)\times(x_2+x_3)$ \citep[cf.][]{bal01},
where $Y_i$ is the (initial) helium abundance in the interior,
$x_2$ and $x_3$ are fractions of He~II and He~III, respectively.

\begin{figure}
\epsfig{file=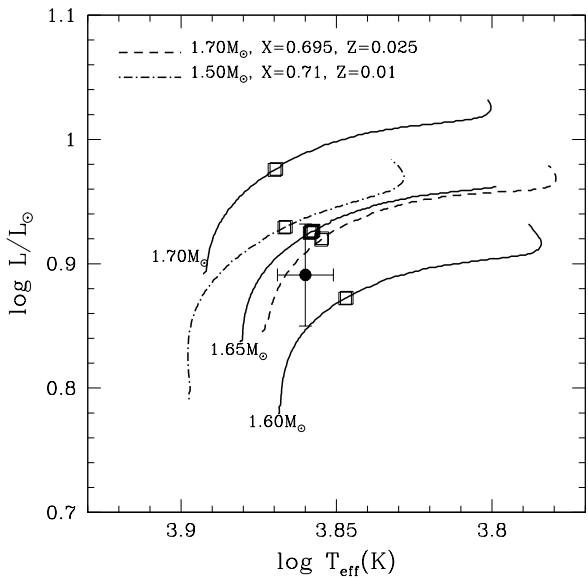,width=0.49\textwidth}
\caption{Evolutionary tracks and the position of HD~24712 with error
bars. Solid lines show the evolutionary tracks for
a standard composition of $(X,Z)=(0.7,0.02)$.
Open squares indicate the loci of models giving best fits to 
the observed frequencies for various cases listed in Table~\ref{tab:models}.
At those positions, models have large frequency separations similar
to 68\,$\mu$Hz, which is determined mainly by stellar mass and radius. 
}
\label{fig:hrd}
\end{figure}

Fig.~\ref{fig:hrd} shows the evolutionary tracks of 
AD160, AD165, AD170, AD170Z25, and AD150Z1  
with the position of HD~24712 with error bars.
The type of $T-\tau$ relation, on/off of the He-depletion, 
and on/off of the envelope convection hardly change the evolutionary
tracks on the HR diagram. 
The stellar masses for the metal rich ($Z=0.025$) and metal poor ($Z=0.01$) 
cases are chosen such that the evolutionary tracks pass close to the
position of HD~24712.
 
We have adopted the effective temperature
$T_{\rm eff}= 7250 \pm 150$\,K (or $\log T_{\rm eff}= 3.860\pm0.009$)
obtained by \citet{ry97} and confirmed by \citet{sh09}.
The range of effective temperature is consistent with 
$T_{\rm eff}=7330\pm 140$\,K derived by \citet{wa97} and
$T_{\rm eff}=7350$\,K determined by \citet{lu08}.
 
\citet{sh09} estimated the radius of HD~24712 to be
$R = 1.772\pm0.043{\rm R}_\odot$ by combining the spectral energy distribution
and the Hipparcos parallax $\pi=20.32\pm0.39$~mas \citep{van07}.    
These estimates for the effective temperature and radius yield
$\log L/{\rm L}_\odot = 0.891\pm 0.041$ for the luminosity of HD~24712, which was 
used to place HD~24712 on the HR~diagram (Fig.~\ref{fig:hrd}).
We note that the adopted luminosity and temperature of HD~24712 are 
consistent with the values used in previous analysis by \citet{cu03} 
($\log L/{\rm L}_\odot = 0.892\pm0.041$ and $\log T_{\rm eff}= 3.869^{+0.006}_{-
0.012}$), 
where the luminosity (based on the Hipparcos parallax) 
was taken from \citet{ma99}.

\subsection{Pulsation Models}
We have obtained nonadiabatic frequencies and eigenfunctions 
for axisymmetric high-order p-modes based on the method
described in \citet{sa05} except for the outer-boundary condition
and the perturbation of radiation.
Since all of the observed oscillation frequencies in HD~24712 are above
the acoustic cut-off frequencies of the models considered here,
we used a running wave condition for the mechanical
outer boundary condition at $\log\tau\approx-6$.

Applying a standard method \citep[see e.g.,][]{unnoetal} to the mechanical
equations (A3) and (A4) of \citet{sa05}, we obtain a running-wave 
condition as
\begin{equation}
{V\over2\chi_\rho}\left[1-i\sqrt{4\omega^2\chi_\rho/V-1}\right]\bmath{Y}_2
= \omega^2\bmath{Y}_1,
\end{equation}
where $\chi_\rho\equiv(\partial\ln P/\partial\ln\rho)_T$ with $P$ and $\rho$ 
being pressure and matter density, respectively, and the other symbols 
are the same as in \citet{sa05}.
In deriving the equation we have assumed the oscillations are 
isothermal at the outer boundary.

In calculating the perturbation of radiation flux,
we have adopted the \citet{un66} theory for the Eddington approximation,
where the radiative flux is given as
\begin{equation}
\bmath{F}_{\rm rad}=-{1\over3\kappa\rho}\nabla\left(acT^4
+{1\over\kappa}{dS\over dt}\right),
\label{eq:edd}
\end{equation}
where $S$, $\kappa$, $a$, and $c$ are entropy per unit mass, 
opacity per unit mass, 
the radiation constant, and the speed of light, respectively. 
We used a linearized form of the above equation \citep{sc80}.
Including the $dS/dt$ term with the running wave boundary condition
is found to reduce the oscillation phase variation in the outermost 
layers making it consistent with observations.

The angular dependences of eigenfunctions
are expanded using axisymmetric spherical harmonics $Y_\ell^{m=0}$ 
(i.e., Legendre polynomials) with $\ell=2j-2$ (even modes) or
$\ell=2j-1$ (odd modes), where $j = 1, 2, \ldots j_t$.
An even (odd) mode is symmetric (anti-symmetric) with respect to
the magnetic equator. 
The expansion is truncated as $j_t=12$ in most cases; but sometimes
$j_t=14$ is adopted to have better accuracy in frequencies.
To label the type of angular dependence of a mode we use $l_m$ which is equal
to the $\ell$ value of the component having the maximum kinetic
energy among the expansion components.
We note that  the angular distribution of amplitude varies depending on 
the strength of magnetic field even for a fixed $l_m$.
To represent the strength of a dipole magnetic field we use the
polar strength $B_p$.

\section{Theoretical results and comparisons to observations}\label{comparison}
\subsection{Oscillation frequencies}\label{subsec:freq}
We calculated nonadiabatic oscillation frequencies between
$\sim2.5$~mHz and $\sim2.9$~mHz for $0 \le l_m \le 3$
including the effect of
dipole magnetic fields in a range of $2\le B_p/({\rm kG}) \le 7$.
For the series listed in Table~\ref{tab:models}, we performed 
pulsation analyses for models having large frequency spacings comparable with 
that of HD~24712; i.e., 68\,$\mu$Hz.

All the pulsation modes having frequencies comparable with the
observed ones are found to be damped; i.e., no excited modes are found.
The excitation mechanism for the oscillations of roAp stars is
generally thought to be the $\kappa$-mechanism in the hydrogen ionization
zone \citep{bal01,cu02}.
For cool roAp stars like HD~24712 and HD 101065 \citep{mkr08}, however,
the $\kappa$-mechanism seems not strong enough.
The temperature inversion (Fig.~\ref{fig:ttau}) is too small
to help excite high-order p-modes as discussed in \citet{gsh98}.
It seems that we need a new excitation mechanism for cool roAp stars.

\begin{figure}
\epsfig{file=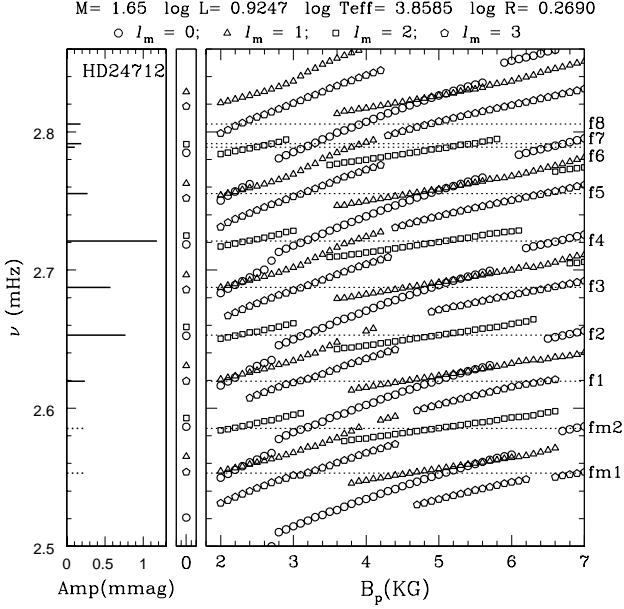,width=0.49\textwidth}
\caption{
Observed oscillation frequencies and amplitudes of HD~24712 (left panel) and 
theoretical frequencies as a function of $B_p$ 
(the strength of magnetic field at poles) for a model 
in the series AD165 (right panel).
In the left panel, solid lines show frequencies and the photometric amplitudes 
obtained by the WET campaign \citep[f1--f8;][]{ku05}, 
while dotted lines represent additional two frequencies (fm1 and fm2) 
obtained by the radial velocity observation by \citet{mkr05}, 
the amplitudes of which are set arbitrary equal to 
the amplitude of the WET lowest frequency.
The label for each frequency is indicated along the rightmost vertical axis.  
The middle panel shows theoretical frequencies obtained without including
the magnetic effects. Horizontal dotted lines in the right panel 
indicate the observed frequencies.  
}
\label{fig:nubp}
\end{figure}

Fig.~\ref{fig:nubp} shows an example of p-mode frequencies in a model as
a function of $B_p$ (the magnetic-field strength at poles). 
The observed frequencies by a WET campaign \citep{ku05} and
the radial velocity observation by \citet{mkr08} are also shown.
Different symbols correspond to different values of $l_m$.
Generally, the frequency of an oscillation mode increases with $B_p$,
but occasionally it jumps down by $\sim30~\mu$Hz and then starts 
increasing again. This property was discovered by \citet{cg00} and 
confirmed by \citet{sg04}. The expansion method adopted in this 
investigation fails around the frequency jump, where the distribution
of kinetic energy among the components of expansion is broad.

\begin{figure}
\begin{center}
\epsfig{file=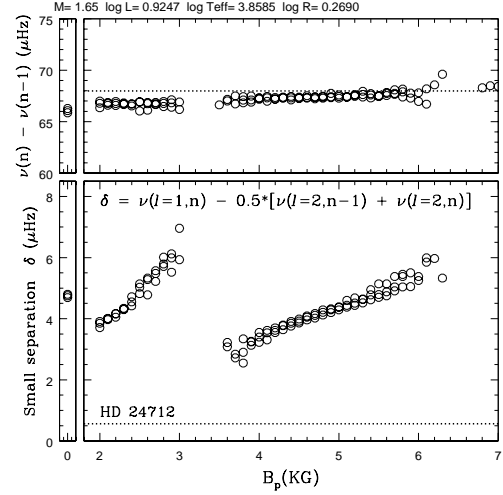,width=0.4\textwidth}
\end{center}
\caption{Large separation (top panel) and small separation (bottom panel)
of frequencies as functions of $B_p$ for the model shown in Fig.~\ref{fig:nubp}.
Dotted lines indicate observed large and small separations of HD~24712.
}
\label{fig:small}
\end{figure}

Frequencies of different radial order for a given $l_m$ vary approximately
parallel as a function of $B_p$ so that
the large frequency separation $\Delta\nu$ hardly changes with $B_p$, 
where $\Delta\nu \equiv \nu(l_m;n)-\nu(l_m;n-1)$ 
with $n$ being the radial order of a mode.
The top panel of Fig.~\ref{fig:small} shows $\Delta\nu$
calculated for $l_m=2$ modes in the model shown in Fig.~\ref{fig:nubp}.
The large separation increases very slowly as a function of $B_p$.
This corresponds to an increase in the phase velocity of magneto-acoustic
wave, $\sqrt{c_s^2+v_A^2}$, with $B_p$, where $c_s$ and $v_A$ are 
the adiabatic sound speed and the Alfv\'en speed, respectively.     

The bottom panel of Fig.~\ref{fig:small} shows the small separation defined as
\begin{equation}
\delta = \nu(l_m=1;n) - 0.5[\nu(l_m=2;n-1)+\nu(l_m=2;n)]
\label{eq:small}
\end{equation}
calculated using the frequencies shown in Fig.~\ref{fig:nubp}.
The small separation of the model is about $5$\,$\mu$Hz at $B_p=0$.
In the presence of a magnetic field, $\delta$ varies from 
$\sim 3$\,$\mu$Hz to $\sim 6$\,$\mu$Hz. 
Generally, $\delta$ increases with $B_p$, but it jumps down
at the frequency jumps discussed above.
The gradual growth of $\delta$ comes from the fact that the frequencies
of $l_m=1$ modes increase slightly more steeply with $B_p$ 
than those of $l_m=2$ modes.   
Just after the jump, $\delta$ reaches a minimum which is considerably
smaller than in the case of $B_p=0$, but still significantly larger than
the observed value $\sim 0.5$\,$\mu$Hz, obtained by assuming that
f3 is a $l_m=1$ mode and f2 and f4 are $l_m=2$ modes.
Such a small value of $\delta$ cannot be reproduced by any
model examined in the present paper.
The problem of the small spacing of HD~24712 is not unique. 
\citet{br09} found a small spacing 
that is essentially zero
in another roAp star $\alpha$ Cir -- 
very much smaller than that of HD~24712.
It is important to solve this problem of the small spacing; 
it could be related to a fundamental property of roAp stars.

A relatively large theoretical $\delta$ means that the frequency 
of an $l_m=1$ mode is slightly larger than the mean of the 
adjacent two $l_m=2$ modes.
It is interesting to note that $l_m=0$ modes cross $l_m=1$ modes
at $B_p\approx 5$~kG (Fig.~\ref{fig:nubp}).
In other words, the frequency of each $l_m=0$ mode is slightly smaller than
the frequency of the adjacent $l_m=1$ mode at $B_p\approx 5$~kG. 
Therefore, the problem of equal spacings of HD~24712 can be solved
if we assign $l_m=0$ modes (rather $l_m=1$ modes) to f1, f3, and f5, 
and consider that $B_p$ happens to have a value  
near the crossings between  $l_m=0$ and $l_m=1$ modes. 
The solution has some weaknesses, however: 1) For the $l_m=(2,0,2)$ 
combination to become equally spaced, $B_p$ should have a particular
value at which the frequency of an $l_m=0$ mode is just equal 
to the mean of the two adjacent $l_m=2$ modes;
it looks unlikely for such a coincidence to be realized in $\alpha$ Cir, too; 
2) As we will discuss below, the amplitude of an $l_m=0$ mode
modulates with rotation phase differently from the observed total 
light variations.  
Nevertheless, we will consider both possibilities for the mode
identifications for f1, f3, and f5.

\begin{figure}
\epsfig{file=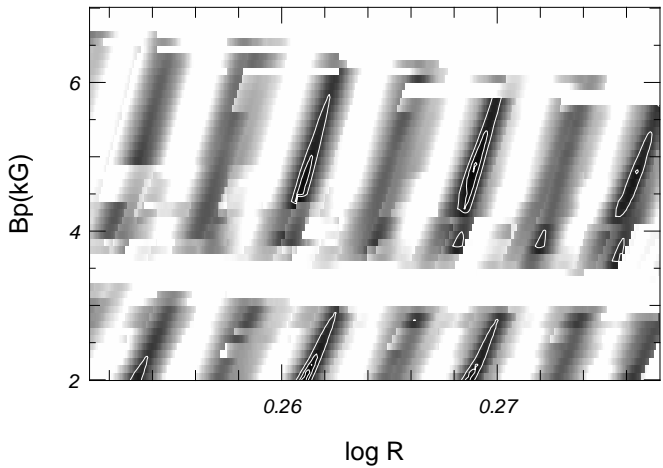,width=0.49\textwidth}
\caption{The distribution of the mean deviation of theoretical frequencies
from the observed ones on the plane of $B_p$ and stellar radius along
the evolution sequence AD165.
Darker parts have smaller mean deviations. Contours drawn for
mean deviations of $1.5$\,$\mu$Hz, $2$\,$\mu$Hz, and $3$\,$\mu$Hz.
The white areas have mean deviations greater than $10$\,$\mu$Hz.
Although the mean deviation in this diagram was calculated without including
$l_m=0$ modes, the distribution of the mean deviation with $l_m=0$ modes
is similar. 
}
\label{fig:meandev}
\end{figure}

The observed frequencies of HD~24712 were compared to models having 
$\Delta\nu\sim68$\,$\mu$Hz  for each evolution sequence listed in 
Table~\ref{tab:models}. 
The quality of the fits was estimated by the mean deviation of 
model frequencies (for $l_m \le 3$) from the nine observed ones. 
The WET f7 frequency
was not included in calculating the mean deviation because it differs
from f6 only by $2.4$\,$\mu$Hz and no models can fit both frequencies.
In calculating mean deviations, we adopted equal weights for all the
frequencies because observational errors are much
smaller than the theoretical uncertainties which can be estimated
as $\sim1$\,$\mu$Hz from the scatter seen in Fig.~\ref{fig:small}.

\begin{table}
\caption{Best fit models}
\label{bestmodels}
\begin{tabular}{@{}lccccc}
\hline
Name & $\log R$ & $\log T_{\rm eff}$ & $\log L$ & $B_p$ & MD($\mu$Hz) \\
\hline
AD160    &  0.2654 & 3.8471 & 0.8720 & 5.3 &  1.54 (1.09)\\
AD165    &  0.2690 & 3.8585 & 0.9247 & 4.9 &  1.45 (1.04)\\
AD170    &  0.2724 & 3.8696 & 0.9759 & 4.6 &  1.60 (1.05)\\
AH165    &  0.2702 & 3.8581 & 0.9255 & 5.5 &  1.45 (1.03)\\
AH165C   &  0.2723 & 3.8574 & 0.9267 & 6.8 &  1.40 (1.06)\\
AD170Z25 &  0.2735 & 3.8549 & 0.9192 & 5.1 &  1.48 (0.99)\\
AH170Z25 &  0.2668 & 3.8572 & 0.9150 & 5.6 &  1.45 (1.43)\\
AD150Z1  &  0.2548 & 3.8667 & 0.9291 & 4.6 &  1.71 (1.22)\\
AH150Z1  &  0.2561 & 3.8663 & 0.9299 & 5.3 &  1.64 (1.12)\\
SD160    &  0.2665 & 3.8467 & 0.8726 & 5.8 &  1.49 (1.06)\\
SD165    &  0.2700 & 3.8582 & 0.9253 & 5.4 &  1.39 (1.02)\\
SD170    &  0.2733 & 3.8693 & 0.9764 & 5.0 &  1.57 (1.15)\\
SH165    &  0.2711 & 3.8578 & 0.9260 & 6.1 &  1.36 (1.03)\\
\hline
\end{tabular}
\end{table}

Table~\ref{bestmodels} lists parameters and the mean deviation (MD)
of the best-fit model for each model series, where MDs in parentheses
refer to the values obtained including $l_m=0$ modes.
Generally, including $l_m=0$ models yields much better agreement. 
The positions of these best models on the HR diagram
are shown by square symbols in Fig.~\ref{fig:hrd}.
The position of HD~24712 on the HR diagram 
is consistent with 1.65\,M$_\odot$ model with normal composition 
and with 1.70\,M$_\odot$ model with a heavy-element abundance of $Z=0.025$. 
The quality of the frequency fit is practically independent  
of the type of $T$--$\tau$ relations,
helium depletion, or the efficiency of convection.
The mean deviations of the best-fit models tend to be smaller 
when the position of the models on the HR diagram is close to the  
position spectroscopically determined for HD~24712.
  
Although \citet{cu03} concluded that a metal-poor abundance was
preferred, Table~\ref{bestmodels} indicates that our metal-poor
cases are no better than the other cases.  
The discrepancy might arise from the fact that their adopted radius 
of HD~24712, $\log (R/{\rm R}_\odot)\approx 0.231$, is somewhat smaller than ours.

The magnetic field strength for each best fit model is determined mainly
by the requirement that the frequency difference between f8 and f6
be equal to the frequency difference $|\nu(n_4+1;l_m=2)-\nu(n_4+1;l_m=3)|$ 
which varies weakly as a function of $B_p$ (Fig.~\ref{fig:nubp}),
where $n_4$ is the radial order for the main frequency f4; i.e.,
$\nu(n_4;l_m=2)=$f4.
(We note that at $B_p=0$, $|\nu(n;\ell=2)-\nu(n;\ell=3)|$ is always much 
larger than the frequency separation between f8 and f6 (or f7).)
Since the magnetic field effect on pulsations is stronger in less dense
atmospheres for a given frequency and $B_p$, the required magnetic field
strength tends to be smaller for more massive or helium-depleted models.
Considering the fact that the spectroscopically determined magnetic 
field strength is $B_p\approx4.4$\,kG (\S~\ref{intro}), a He-depleted model of 
AD165 or AD170Z25 is better for the model of HD~24712.  

Fig.~\ref{fig:meandev} shows the distribution of the mean deviations
on the $\log R-B_p$ plane for AD165 models, where $R$ means stellar radius.
Darker parts indicate smaller mean deviations. 
Theoretical frequencies are interpolated with respect to $\log R$ along
the evolutionary model sequence, but not interpolated with respect to $B_p$
(frequencies are obtained at every $0.1$\,kG).

As stellar radius changes, the oscillation frequencies change with keeping 
$\Delta\nu$ approximately constant. 
Therefore, at a fixed $B_p$ the mean deviation becomes small cyclically 
with varying $\log R$, when regularly spaced observed frequencies 
(fm1, fm2, and from f1 to f6) become close to $l_m=1$ and $2$ modes.
This corresponds to the cyclic appearances
of dark bands in Fig.~\ref{fig:meandev}.

The dark bands are inclined because oscillation frequencies increase
gradually as $B_p$ increases. They are interrupted at $B_p\sim 3.5 - 3$\,kG,
because oscillation frequencies jump there (see Fig.~\ref{fig:nubp}).
The darkness of the bands varies alternatively; i.e., the mean deviation
is smaller when the frequencies fm2, f2, f4 and f6 are fitted with $l_m=2$ 
modes (as at $B_p\approx 4.9$\,kG in Fig.~\ref{fig:nubp})
rather than with $l_m=1$ modes, because in the former case the frequency
f8 can be fitted well with an $l_m=3$ mode.  

Thus, we identify the main frequency f4 as an $l_m=2$ mode because then
the highest frequency f8 can be well fitted with an $l_m=3$ mode.
This identification is different from that of \citet{ku05}; they identified
f4 as a deformed dipole ($l_m=1$) mode mainly based on the consistency 
of the amplitude of rotational side-lobes. 
We will show below that due to a strong deformation of the amplitude
over the stellar surface, the mean amplitude of the rotational side-lobes
for f4 agrees with that expected from a $l_m=2$ mode.

\begin{figure}
\epsfig{file=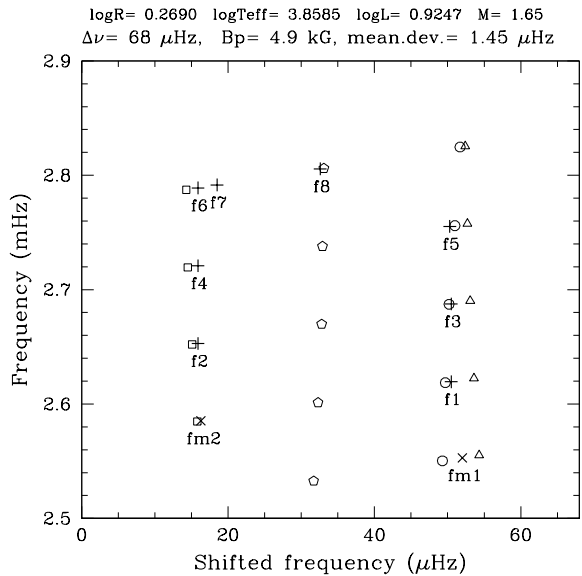,width=0.49\textwidth}
\caption{Echelle diagram of frequencies for the best fit model in AD165
and observed frequencies of HD~24712 from the WET campaign 
\citep[pluses;][]{ku05} and from \citet{mkr05}.
Open symbols are model frequencies; circles for $l_m=0$, triangles for
$l_m=1$, squares for $l_m=2$ and pentagons for $l_m=3$. 
}
\label{fig:echelle}
\end{figure}

Fig.~\ref{fig:echelle} is an Echelle diagram for the model at
$(\log R,B_p) = (0.269, 4.9\rm{kG})$ in Fig.~\ref{fig:meandev}.
It is the best from the family of AD165 models for $B_p > 3$\,kG with
the mean deviation of the model being $1.45$\,$\mu$Hz without
$l_m=0$ modes and $1.04$\,$\mu$Hz if $l_m=0$ modes are included.
(Although a smaller mean deviation of $1.21$\,$\mu$Hz is realized at 
$(\log R,B_p) = (0.261, 2.1\rm{kG})$, the magnetic field strength in the model
is smaller than the observational estimates (\S1), and the phase 
variation in the atmosphere is inconsistent with observation.)
Fig.~\ref{fig:echelle} shows that although the large separation of the model
agrees very well with observations, the small spacing defined by
$l_m=1$ and $l_m=2$ modes is too large by about $3$\,$\mu$Hz. 
This discrepancy disappears if f1, f3 and f5 are fitted with $l_m=0$
modes rather than $l_m=1$ modes.
In this case, the nearly equal spacings, 
$f5-f4\approx f4-f3\approx f3-f2\approx f2-f1 \approx 34$\,$\mu$Hz,
are realized by pairs of $l_m=2$ and $l_m=0$.

In previous investigations the identity of f8 was puzzling because
it is separated from f7 (or f6) by only 17\,$\mu$Hz, 
half of the regular spacing.
In our models $f8$ can be well fitted with an $l_m=3$ mode. 
It is not clear, however, why no other $l_m=3$ modes are
detected.

\subsection{Amplitude modulation with rotation phase}\label{modulation}

\begin{figure}
\epsfig{file=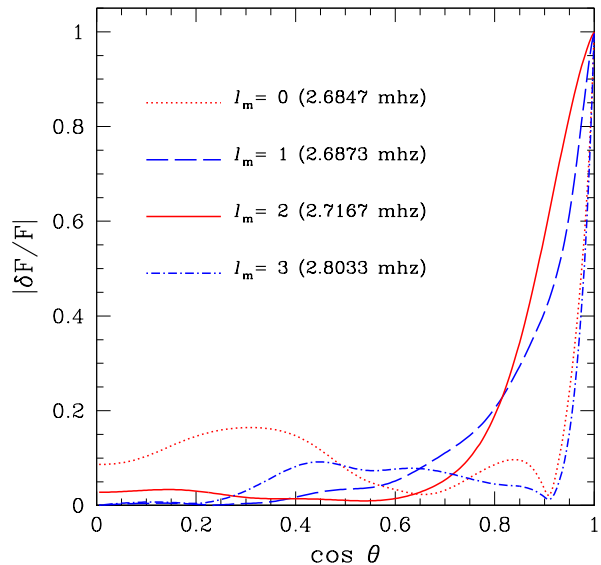,width=0.49\textwidth}
\caption{
The amplitude of radiative flux variation on the stellar surface 
as a function of $\cos\theta$
for a model close to the best fit model of AD165,
where $\theta$ is the co-latitude measured from the magnetic axis.
We have chosen one mode for each degree $l_m$ having a frequency 
close to one of the observed frequencies of HD~24712.
}
\label{fig:theta_amp}
\end{figure}

\begin{figure}
\epsfig{file=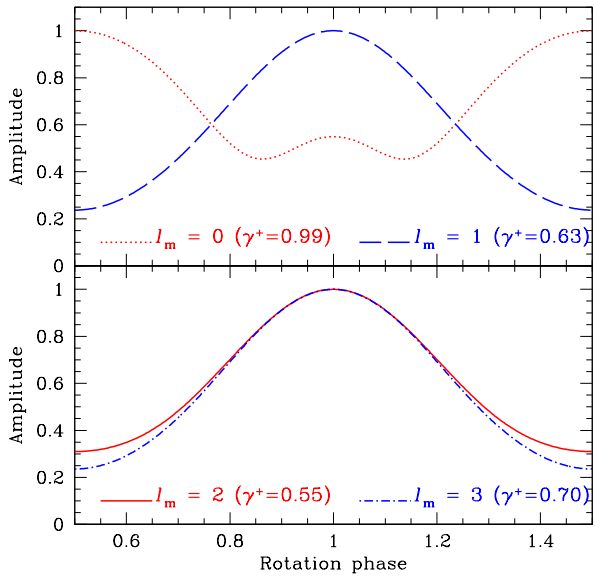,width=0.49\textwidth}
\caption{
Amplitude modulations with rotation phase in light variations 
predicted from the amplitude distributions
on the stellar surface shown in Fig.~\ref{fig:theta_amp}.
We have assumed $i=137^{\circ}$ and $\beta=150^\circ$ according to
\citet{bag95}, and $\mu = 0.6$ for the limb-darkening parameter. 
The quantity $\gamma^+$ is defined as $(A_{-1}+A_{+1})/A_0$ with
$A_0$ and $A_{\pm1}$ being central amplitude and amplitudes of
rotational side-lobes, respectively.
}
\label{fig:ampmod}
\end{figure}

Observed pulsation amplitudes of roAp stars change with magnetic 
(rotation) phase in such a way that the amplitude is maximum 
at the phase of magnetic maximum. 
This is interpreted by the oblique pulsator model proposed by \citet{ku82};
i.e., an axisymmetric pulsation mode whose axis aligns with
the magnetic axis which is in turn inclined to the rotation axis.
As the star rotates one observes pulsations at varying aspect, which 
causes the amplitude modulation.

HD~24712, like other roAp stars, shows an amplitude modulation with the
amplitude maximum occurring at the magnetic maximum as
found by \citet{ku89} \citep[see also][]{pap1}. 
Theoretically, amplitude modulation of a pulsation mode  can be obtained 
by integrating the amplitude distribution over the stellar
disc assuming various values of the angle between pulsation axis 
(i.e., magnetic axis) and the line-of-sight expected during a rotation
period. 

Fig.~\ref{fig:theta_amp} shows the amplitude of radiative flux variation
as a function of $\cos\theta$ for individual pulsation modes with 
various latitudinal degrees $l_m$ for
the best model of AD165. $\theta$ is the co-latitude
with respect to the magnetic axis.
The frequency of each mode is close to one of the observed frequencies of 
HD~24712, though the amplitude distribution is insensitive to the pulsation
frequency.
Due to the presence of a strong magnetic field, the latitudinal distribution
of pulsation amplitude significantly deviates from any single Legendre
function $P_\ell(\cos\theta)$.
The amplitude around the equatorial region is strongly suppressed and 
it is more concentrated in the polar regions compared to 
the non-magnetic case.
The $l_m=0$ case is somewhat different from the other cases; it has 
a broad peak around the equator, which significantly influences
amplitude modulation as discussed below.   

It is remarkable that the amplitude distribution of the $l_m=1$ mode 
in the hemisphere is very close to that of the $l_m=2$ mode except
that the latter (former) is symmetric (anti-symmetric) to the 
equatorial plane.
Since we see mostly one magnetic hemisphere of HD~24712,
we expect that amplitudes in light variations 
and rotational amplitude modulations of 
$l_m=1$ and $l_m=2$ modes are comparable to each other.  

Fig.~\ref{fig:ampmod} shows amplitude modulations predicted from the
amplitude distributions shown in Fig.~\ref{fig:theta_amp}, assuming
Bagnulo et al.'s (1995) parameters: $i=137^\circ, \beta=150^\circ$. 
Magnetic maximum corresponds to the rotation phase of unity.
All the modes except for $l_m = 0$ have maximum amplitudes at magnetic
maximum in agreement with the observations by \citet{ku89}.

Despite a considerable difference in the amplitude distributions
between $l_m=1$ and $l_m=3$ (Fig.~\ref{fig:theta_amp}), 
the amplitude modulation curves are very similar. 
This comes from the fact that for an odd mode contributions from components 
of $\ell\ge3$ to the light variation is very small compared with that of 
the $\ell=1$ component, as discussed in \citet{sg04}.

In contrast to the other cases, amplitude of the $l_m=0$ mode is maximum at
the magnetic minimum, although the modulation amplitude is small.
This is caused by a broad peak of amplitude near the equator seen 
in Fig.~\ref{fig:theta_amp}.
The property of the amplitude modulation of the $l_m = 0$ mode indicates 
that the largest amplitude mode of HD~24712 cannot be matched with
a $l_m = 0$ mode, although it is still possible that
$l_m=0$ modes are involved in the pulsations of this star without
influencing the total signal very much.  

The observed ratio of the minimum to the maximum amplitude can be read as 
$\sim 0.3$ from Fig.~1 of \citet{ku89}, which is consistent with 
the theoretical predictions for $l_m= 1, 2$ and 3.
More quantitative comparisons are possible by using rotational side-lobes
in the Fourier spectrum.
The rotational side-lobes are characterized by the quantity
\begin{equation}
\gamma^+ \equiv {A_{-1}+A_{+1}\over A_0} ,
\end{equation} 
where $A_0$ refers to the amplitude of the central frequency and
$A_{-1}$ and $A_{+1}$ refer to the amplitudes of the first 
lower- and higher-frequency side-lobes, respectively.
(For our theoretical side-lobes $A_{-1}=A_{+1}$ because we do not include 
Coriolis force effects.)  
The value of $\gamma^+$ for each mode in Fig.~\ref{fig:ampmod}
is indicated in parentheses.  The value of $\gamma^+$ depends on
the latitudinal degree $l_m$, but it 
is insensitive to the frequency; i.e., pulsation modes with the same $l_m$
but different radial orders have similar values of $\gamma^+$.

\citet{ku05} obtained $\gamma^+$ for each of the 
frequencies of HD~24712;
among them the values for the three main frequencies f2, f3 and f4
are well determined with uncertainties less than 6\%.
The mean value from 2000 and 1986 data is 0.60 for f4, and 
0.64 for f2. We have matched f4 and f2 with $l_m=2$ modes
which have $\gamma^+=0.55$ (Fig.~\ref{fig:ampmod}), not very
different from the observed values.
On the other hand, the mean value for f3 is 0.83, which is considerably
higher than 0.63 expected for an $l_m=1$ mode. 
As discussed above and indicated in Fig.~\ref{fig:echelle}, 
f3 is also close to an $l_m=0$ mode having $\gamma^+=0.99$.
The observed value of $\gamma^+$ lies in the middle
between $\gamma^+$s of $l_m=1$ and $l_m=0$ modes.
This could indicate the possibility for the frequency f3 to be a superposition 
of the $l_m=1$ and $l_m=0$ modes.

The amplitudes of the second rotational side-lobes are predicted to be
about 10 times smaller than those of the first side-lobes.
For example, the amplitude of the second side-lobes of f4 ($l_m=2$)
is expected to be about 30\,$\mu$mag, which is smaller than
the amplitude of the highest noise peaks ($80$\,$\mu$mag) 
in the analysis of \citet{ku05}.

\begin{figure}
\epsfig{file=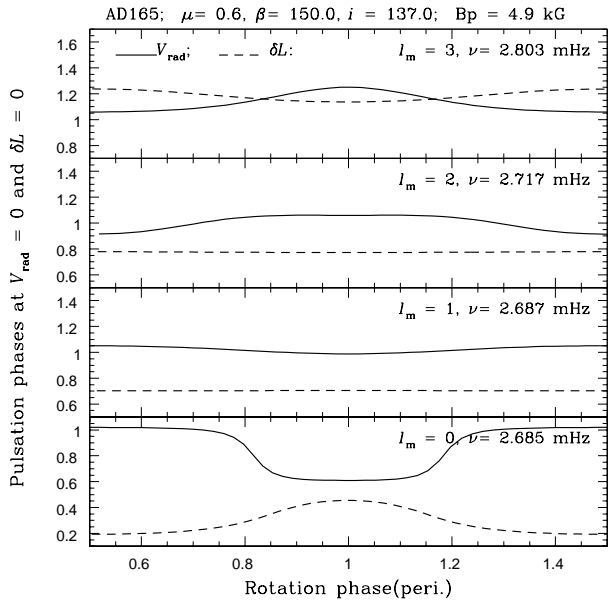,width=0.49\textwidth}
\caption{
Theoretical oscillation phase modulations in light and velocity variations
for each mode shown in Fig.~\ref{fig:theta_amp}. 
In each panel oscillation phases at $\delta L= 0$ and at $V_{\rm rad}=0$
(from negative to positive) are plotted as a function of the
rotation phase, where $V_{\rm rad}$ refers to the radial velocity
at $\log\tau\approx -6$.
}
\label{fig:phasemod}
\end{figure}

Fig.~\ref{fig:phasemod} shows rotational modulations of pulsation phase
at  $\delta L= 0$ and at $V_{\rm rad}=0$ (changing from negative to positive) 
for each mode shown in Fig.~\ref{fig:theta_amp},
where $V_{\rm rad}$ is the radial velocity (RV) at $\log\tau\approx-6$. 
Except for the $l_m=0$ case no rapid changes occur in pulsation phase as
observed in some other stars \citep[e.g. HR~3831 (HD~83368),][]{ks86,ba00}. 
This is  because we see only one magnetic pole of HD~24712 with 
a configuration of $(\beta,i)=(150^\circ,137^\circ)$. 
For the $l_m=0$ mode rapid changes in pulsation phase  occur at
the rotation phases  $\sim0.8$ and $\sim1.2$. 
This comes from the presence of a nodal line at $\cos\theta \approx 0.9$
in the amplitude distribution (Fig.~\ref{fig:theta_amp}). 

For the $l_m=2$ mode (the 2nd panel from the top in Fig.~\ref{fig:phasemod}), 
which corresponds to the main frequency f4, the phase of the luminosity variations
is nearly constant. 
The phase of the RV variations is also practically constant in the rotational 
phase interval $0.75 - 1.25$, although it changes gradually at other phases. 
Simultaneous spectroscopic and photometric monitoring of 
HD~24712 \citep{SRB06} (see also Paper I) performed between rotational
phases 0.87 and 1.18 do not show phase variations, 
thus supporting theoretical predictions.   
The phase of the luminosity variations is 
always smaller than that of the RV variations, 
which means that the maxima (for example) of the luminosity variation precede 
the maxima of the RV variations.
The maximum value of the phase difference is about 0.3 of the period 
and is realized when the rotation phase is around unity; 
i.e., when the pulsation amplitude is maximum (Fig.~\ref{fig:ampmod}).
This means that velocity maxima lag luminosity maxima by 
about 30\% of a pulsation period around the rotation phase 
when oscillation amplitude is maximum.

\begin{table}
\caption{Phase lags in the outermost layers}
\label{phaselag}
\begin{tabular}{@{}lccc}
\hline
Line & $\lambda$(\AA) &$\log\tau_{5000}$& $\delta\phi$ \\
\hline
 Pr III & 5285 &  -6.41 &  0.44 \\
 Pr III & 5300 &  -6.35 &  0.46 \\
 Pr III & 6160 &  -6.33 &  0.42 \\ 
 Pr III & 6090 &  -6.30 &  0.43 \\ 
 Pr III & 6196 &  -6.29 &  0.43 \\ 
 Nd III & 5294 &  -6.03 &  0.33 \\ 
 Nd III & 5127 &  -5.94 &  0.31 \\ 
 Nd III & 4927 &  -5.79 &  0.28 \\ 
 Pr III & 6707 &  -5.53 &  0.37 \\
 Pr III & 5844 &  -5.51 &  0.38 \\
 Pr III & 5999 &  -5.50 &  0.38 \\
 Nd III & 6145 &  -5.49 &  0.28 \\
 Pr III & 6053 &  -5.48 &  0.39 \\
 Nd III & 5410 &  -5.44 &  0.26 \\
 Nd III & 5988 &  -5.43 &  0.27 \\
 Nd III & 5845 &  -5.32 &  0.27 \\
\hline
\end{tabular}
\end{table}

\citet{pap1} obtained phase lags of RV variations 
with respect to the light variations for various spectral lines.
These phase lags are always negative which means that luminosity 
maximum occurs after the RV maxima. 
For two other roAp stars, 10~Aql \citep{SKR08} and HD~101065 \citep{mkr08}, 
the observed phase lags are positive, 
in the range $0.4-0.6$ (10~Aql) and $0.16-0.19$ (HD~101065), respectively.  
Allowing one cycle for the phase uncertainty, the negative phase lags 
in HD~24712 may be converted into the positive ones.
Table~\ref{phaselag} lists some results for the lines formed
in the most superficial layers, where optical depth of the
formation of each line has been obtained by using a NLTE model of \citet{sh09}.
Generally the phase lag $\delta\phi$ decreases with depth, indicating
oscillations propagates outward as discussed below.
Table~\ref{phaselag} shows that the phase lag around $\log\tau\approx-6$
is about $0.3-0.4$, roughly consistent with the theoretical 
prediction for the main frequency.

\subsection{Amplitude and phase variations with depth}\label{depth}

\citet{pap1} obtained amplitudes and phases of the RV variations 
in various spectral lines for the two highest amplitude modes (f4 and f2) 
from time-resolved spectroscopic observations of HD~24712. 
In the present paper we consider the results at a rotational phase of 0.94 
(UVES observations close to the magnetic maximum). 
For a representative sample of \pra, \prb, \nda and \ndb\ lines, as well as 
for the \ha\, core, depths of formation were calculated in NLTE approximation 
\citep{MRR05,mash09} using a model atmosphere from \citet{sh09}. 
In these iterative model calculations empirical stratifications of 
the chemical elements Si, Ca, Cr, Fe, Sr, Ba, Pr and Nd 
were derived for each iteration. 
NLTE calculations were performed for Pr and Nd only, 
because these elements are concentrated in the uppermost atmospheric layers, 
while other elements have a tendency to be accumulated close to photosphere 
where NLTE effects may be neglected. 
Final element distributions were then used for depth of formation calculations. 
The results have been converted to the relations between optical depth 
and RV amplitude/phase.
For \ya\, lines no stratification analysis was performed, therefore, depth of 
line formation was calculated with a uniform yttrium
distribution using an yttrium abundance $\log(Y/N_{\rm tot})$=8.60 derived 
from abundance analysis. 

\begin{figure}
\epsfig{file=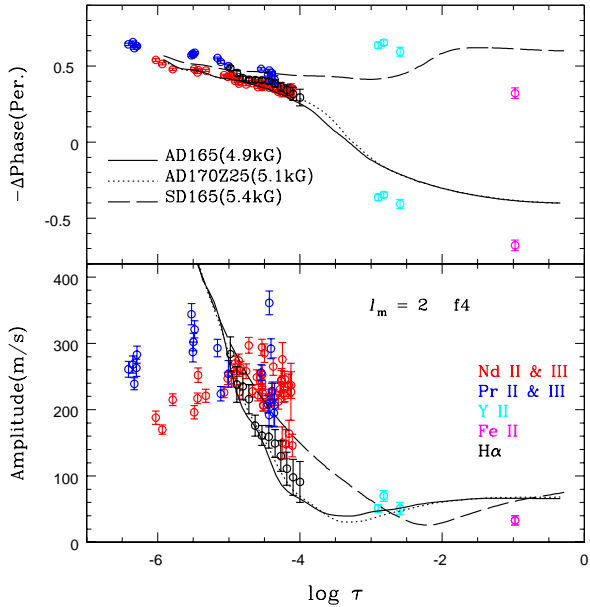,width=0.49\textwidth}
\caption{
Optical depth versus phase delay (top panel) and amplitude
(bottom panel) of RV variations for f4 (the main
oscillation mode) obtained from various spectral lines of HD~24712.
For \ya\ and \fea\ the phases shifted downward by a pulsation period
are also shown.  
Various lines are theoretical relations for $l_m=2$ modes
having frequencies similar to f4  for some He-depleted models
(see Table~\ref{tab:models} for parameters of each model). 
In calculating theoretical relations angles of 
$(i,\beta)=(137^\circ,150^\circ)$ are assumed.
}
\label{fig:tauampD}
\end{figure}

\begin{figure}
\epsfig{file=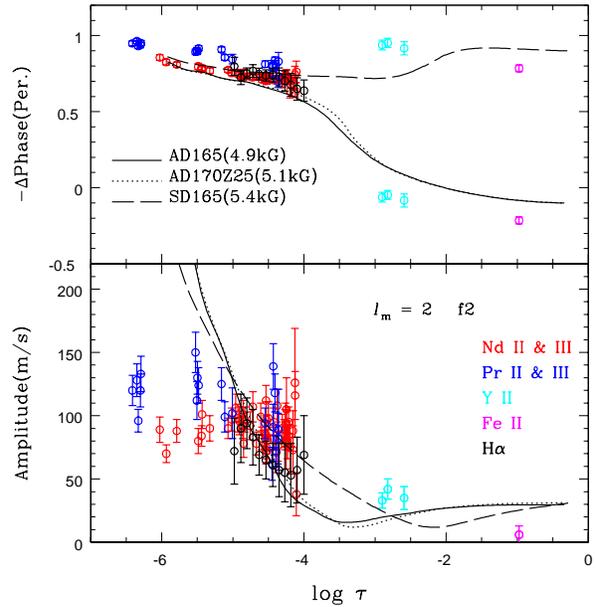,width=0.49\textwidth}
\caption{
The same as Fig.~\ref{fig:tauampD} but for f2.
}
\label{fig:tauampD2nd}
\end{figure}

\begin{figure}
\epsfig{file=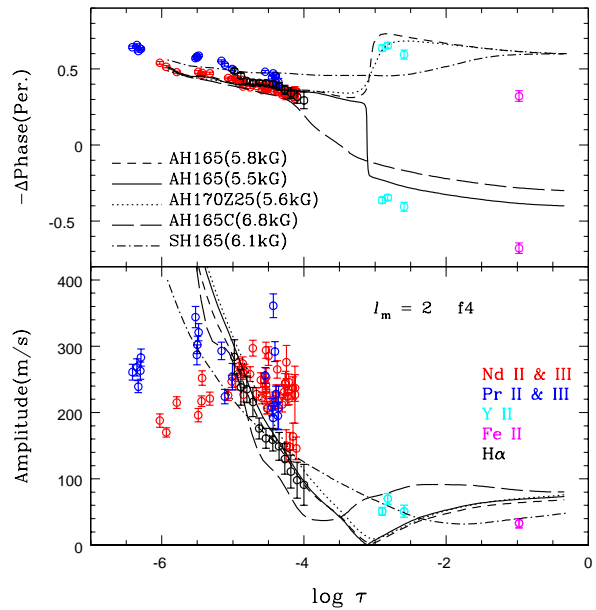,width=0.49\textwidth}
\caption{The same as Fig.~\ref{fig:tauampD} but compared with models
without helium depletion.
}
\label{fig:tauampH}
\end{figure}

\begin{figure}
\epsfig{file=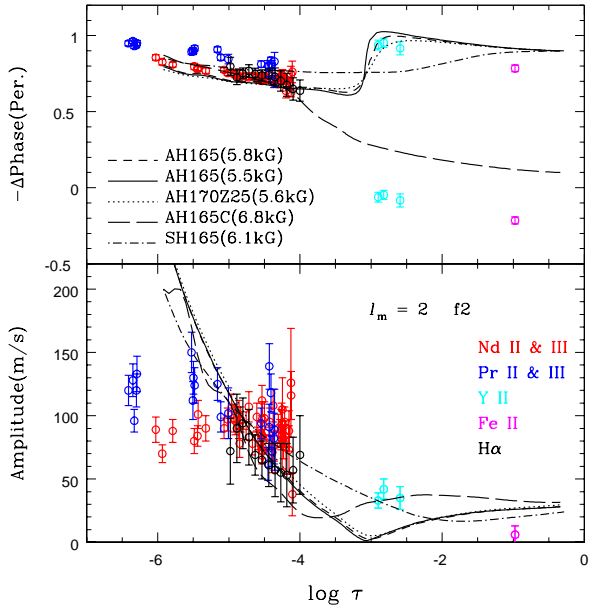,width=0.49\textwidth}
\caption{The same as Fig.~\ref{fig:tauampH} but for f2.
}
\label{fig:tauampH2nd}
\end{figure}

Figs\,\ref{fig:tauampD} and \ref{fig:tauampD2nd} compare RV 
amplitude/phase -- $\log\tau_{5000}$ relations
for f4 and f2, respectively, with the corresponding theoretical relations of 
$l_m=2$ modes for He-depleted best models (Table~\ref{bestmodels}).
For \ya\ and \fea\ lines with very low RV amplitudes pulsational phases 
are plotted twice: 
as defined originally in \citet{pap1}, and shifted downward by 
a pulsation period taking into account the one-period
uncertainty in the phase relation.
Similar comparisons for models without He-depletion are given 
in Figs\,\ref{fig:tauampH} and \ref{fig:tauampH2nd}.  
In comparing the observed and theoretical relations, the theoretical amplitude
is multiplied by an arbitrary factor, and the theoretical phase delay is 
fitted at the outer boundary ($\log ~\tau \approx -6$) by applying
an arbitrary shift.

In the outermost layers oscillation phase gradually delays
toward the outer boundary, indicating the presence of running waves
propagating outward.
This is consistent with the fact that
all the observed oscillation frequencies of HD~24712 are
larger than the acoustic cut-off frequencies of theoretical models.
(If a reflective outer boundary condition is
imposed, the phase in the outermost layers is nearly constant.)
The gradient of the phase variation in the outermost layers
is well reproduced by theoretical relations.
We note that if the $dS/dt$ term in equation~(\ref{eq:edd}) is dropped,
the gradient becomes steeper, indicating that the \citet{un66} theory
is consistent with the observations.

The phase distribution in HD~24712 has a jump
between $\tau\approx10^{-4}$ and $10^{-3}$ for both f4 and f2,
indicating the presence of a quasi-node there. 
This property agrees with the theoretical phase variations for A-models 
with the $T-\tau$ relation having a small temperature inversion 
at $\log\tau\approx-3.5$ (Fig.~\ref{fig:ttau}).
A rapid phase change in S-models with the standard $T-\tau$ relation, however,
occurs between $\tau\approx10^{-3}$ and $10^{-2}$, deeper than the phase
jump derived from observations.
Thus, the position of the phase jump supports the $T-\tau$ relation
with a temperature inversion at $\log\tau\approx-3.5$ as obtained
by \citet{sh09}.

Whether the phase of theoretical models jump downward or upward 
depends on the assumption of helium depletion, and sometimes on
the strength of the magnetic field.
Generally, phase tends to decrease inward at the quasi-node in 
helium-depleted models, and to increase inward in models without helium
depletion. 
In the best model of the AH165 series with $B_p=5.5$\,kG, 
the phase of RV variations 
for the main frequency f4 decreases steeply inward 
at $\log\tau\approx-3$, while it {\it increases} steeply at the same layer 
if $B_p$ is slightly increased to $5.8$\,kG (Fig.~\ref{fig:tauampH}),
although for the secondary frequency f2 the phase increases steeply inward
for both cases (Fig.~\ref{fig:tauampH2nd}). 
However, such a strong dependence of the direction of the phase jump
on $B_p$ does not occur in the other cases. 

To obtain information on the helium depletion in the atmosphere of HD~24712,
further observations are needed to fill the gap 
between $\log\tau=-4$ and $-3$ .

Towards deep interior, theoretical pulsation
phase approaches a constant; i.e., nearly a standing wave is 
realized in the deep atmosphere.
However, the innermost Fe II data for the main
pulsation f4 deviate considerably from the theoretical curves.
A possible cause may be related to the surface element distribution;
Fe is concentrated around the equator, while Y 
(as all rare earth elements, including Pr and Nd) 
is concentrated near magnetic poles \citep{lu08}.

Oscillation amplitude increases rapidly in the outermost layers
due to a rapid decrease in the gas density.
In the layers of $-5 < \log~\tau < -4$, theoretical amplitude
variations roughly agree with the observed ones for both f4 and f2.
In the outermost layers with $\log~\tau < -5$, however, theoretical
amplitudes deviate significantly from the observations.
Observed amplitudes level off in the most superficial layers,
the cause of which is not clear; it could be the effect of nonlinear 
dissipation, or the density stratification could differ significantly
from our simple models.

\section{Conclusions}\label{conclusions}

We discussed theoretical models for the oscillations of the roAp star
HD~24712 (HR~1217). 
Observed frequencies are fitted well with theoretical ones for models
whose positions on the HR diagram are close to the position
of HD~24712 determined by spectroscopy along with the Hipparcos parallax.
The observed main frequency f4 is identified as a quasi-quadrupole 
($l_m=2$) mode, whose amplitude on the surface is strongly concentrated 
in the polar regions and is suppressed significantly around the equator.
The theoretical amplitude distribution predicts a rotational modulation of the 
pulsation amplitude which is consistent with the observations.

We modelled for the first time the distributions of the phase and amplitude 
of RV variations as a function of atmospheric height and compared these 
with the observed distributions.
The gradual outward increase of phase lag in the outermost layers 
is well reproduced by theoretical results obtained with a running-wave 
outer boundary condition.
The presence of a steep phase change between $\log\tau\approx -4$ and $-3$ 
favours a $T-\tau$ relation with a temperature inversion 
at $\log\approx-3.5$, rather than a standard $T-\tau$ relation.

Although our models agree with most of the observed properties of 
the oscillations in the atmosphere of HD~24712, we have recognized 
three problems to be solved in the future.
\begin{enumerate}
\item The observed amplitude of velocity variation levels off in the outermost
layers with $\log\tau < -5$, while in all the models we have calculated
amplitude increases steeply outward in the layers with $\log\tau < -4$.  
\item If the observed main frequencies are identified with $l_m=2$ and $l_m=1$
modes, to explain their nearly equal spacings 
the small frequency spacing defined as
$\nu(n,l_m=1)-0.5[\nu(n,l_m=2)+\nu(n-1,l_m=2)]$ must be as small as 
$0.5$\,$\mu$Hz. 
For all the models calculated, however,
the small spacing was always larger than $\sim3$\,$\mu$Hz. 
The problem goes away if we assume the observed equal spacings of frequencies
are produced by $l_m=2$ and $l_m=0$ modes rather than $l_m=2$ and $l_m=1$ modes.
However, this solution has a problem; 
$l_m=0$ modes have rotational amplitude modulations
different from the modulation seen in the light variations of HD~24712.
\item
We have found no excited oscillation modes with frequencies appropriate 
for HD~24712; in other words all the modes examined are damped ones. 
High-order p-modes in roAp stars are generally thought to be excited
by the $\kappa$-mechanism in the hydrogen ionization zone \citep{bal01,cu02}.
However, the $\kappa$-mechanism does not seem to be strong enough to excite 
the super-critical high-order p-modes in HD~24712.
It is interesting to note that the inefficiency of the $\kappa$-mechanism
is in common with the case of another cool roAp star HD 101065 having 
many regularly spaced frequencies \citep{mkr08}.  
Probably we need to find a new excitation mechanism for these
coolest roAp stars.
\end{enumerate}

\section*{Acknowledgments}
We thank Dr. L. Mashonkina for providing us with the numerical data 
on the NLTE Pr-Nd line depth formation.  
HS thanks Chris Cameron for stimulating discussions.
This work was supported by research grants from the RFBI (08-02-00469a
and 09-02-00002a) and from Leading Scientific School (4354.2008.2) to TR and MS.

\label{lastpage}
\end{document}